\documentclass[aps, prl, amsfonts, amssymb, amsmath, floatfix, superscriptaddress, showpacs, reprint]{revtex4-1}

\usepackage{times}
\usepackage{epsfig}
\usepackage{graphicx}
\usepackage{epstopdf}
\usepackage{bm}
\usepackage{color}
\usepackage[english]{babel}

\PassOptionsToPackage{hyphens}{url}
\usepackage[breaklinks=true,colorlinks=true,citecolor=blue,urlcolor=blue,linkcolor=blue]{hyperref}

\renewcommand{\vec}[1]{\mathbf{#1}}

\begin{document}

\title{Fully Kinetic Simulation of 3D Kinetic Alfv\' en Turbulence}

\author{Daniel~Gro\v selj}
\affiliation{Max-Planck-Institut f\" ur Plasmaphysik, Boltzmannstra\ss e 2, 
D-85748 Garching, Germany}
\author{Alfred~Mallet}
\affiliation{Space Science Center, University of New Hampshire, Durham, NH 03824, USA}
\author{Nuno~F.~Loureiro}
\affiliation{Plasma Science and Fusion Center, 
Massachusetts Institute of Technology, Cambridge, MA 02139, USA}
\author{Frank~Jenko}
\affiliation{Max-Planck-Institut f\" ur Plasmaphysik, Boltzmannstra\ss e 2, 
D-85748 Garching, Germany}

\date{\today}

\begin{abstract}
We present results from a three-dimensional particle-in-cell simulation of plasma turbulence, resembling 
the plasma conditions found at kinetic scales of the solar wind. 
The spectral properties of the turbulence in the subion range are consistent with theoretical 
expectations for kinetic Alfv\' en waves. Furthermore, we calculate the local anisotropy, defined by the relation 
$k_{\parallel}(k_{\perp})$, where $k_{\parallel}$ is a characteristic wave number along the local mean magnetic 
field at perpendicular scale $l_{\perp}\sim 1/k_{\perp}$. 
The subion range anisotropy is scale dependent with $k_{\parallel}<k_{\perp}$ and 
the ratio of linear to nonlinear time scales is of order unity, suggesting that the kinetic cascade 
is close to a state of critical balance. Our results compare favorably against a number of \emph{in situ} solar 
wind observations and demonstrate---from first principles---the 
feasibility of plasma turbulence models based on a critically balanced cascade of kinetic Alfv\' en waves.
\end{abstract}


\maketitle

{\em Introduction.---}Many space and astrophysical plasmas are found in a weakly collisional turbulent state, 
with prominent examples ranging from the solar wind \cite{bruno2013}, 
to more distant astrophysical environments such as 
accretion disks \cite{Quataert1999, Sharma2007, Kunz2016}, 
galaxy clusters \cite{Dennis2005, Zhuravleva2014}, and 
the interstellar medium \cite{Armstrong1995, Lithwick2001}. In low-collisionality plasmas, 
the fluidlike inertial range energy cascade transitions into kinetic turbulence at 
the ion kinetic scales, with important implications for the turbulent heating of ions and electrons, 
and for the (bulk) transport properties of the plasma~\cite{Quataert1999, Cranmer2003, 
Howes2008, Schekochihin2009, Chandran2011, Matthaeus2015, BanonNavarro2016}. 
The nature of the kinetic-scale plasma turbulence 
is, however, still a matter of debate~\cite{Howes2008a, Matthaeus2008, Howes2008, Shaikh2009, Schekochihin2009, 
Podesta2010, Podesta2012, Chen2013, Servidio2015, Wan2015, Howes2015a, Narita2016, Perrone2016}. 
The most detailed observational data originate from \emph{in situ} solar 
wind measurements~\cite{Leamon1998, Bale2005, Alexandrova2009, Sahraoui2010, 
Osman2011, Salem2012, Chen2013, Chasapis2015, Perrone2016, Narita2016}, which thus 
provide the most stringent constraints for the theoretical predictions~\citep{Galtier2003, Galtier2006, Howes2008, 
Schekochihin2009, Boldyrev2013, Passot2015}.
Spacecraft measurements have shown that the solar wind is highly turbulent, displaying 
power-law fluctuation spectra over a broad range of scales \cite{bruno2013, Kiyani2015, Chen2016}.
In the inertial range, above the proton kinetic scales, the magnetic energy 
follows an $E(k_{\perp})\propto k_{\perp}^{-5/3}$ wave number spectrum 
in directions perpendicular to the local mean magnetic field, whereas the inferred spectrum
parallel to the local mean field is steeper: $E(k_{\parallel})\propto k_{\parallel}^{-2}$ 
\cite{Horbury2008, Podesta2009, Wicks2010}. Thus, solar wind turbulence is anisotropic. At 
kinetic scales, a break in the inertial range spectrum is observed, followed by a steeper 
power law with a spectral exponent around $-2.8$ at subproton scales \cite{Alexandrova2009, Sahraoui2010} for wave numbers 
nearly perpendicular to the mean field. Turbulence at kinetic scales remains anisotropic~\cite{Chen2010, Sahraoui2010}, 
although presently available measurements limit the accuracy to which one can determine 
the kinetic-scale anisotropy.

An elegant explanation for the development of scale-dependent anisotropy can be given in terms of the 
critical balance conjecture \cite{Goldreich1995, Cho2000, Maron2001, 
Howes2008, Schekochihin2009, TenBarge2012, Boldyrev2013, Mallet2015, Chen2016}. This states that 
even when the turbulent plasma dynamics is strongly nonlinear, certain properties of linear 
wave physics are maintained, such that the nonlinear time at each scale is 
comparable to the characteristic time of the relevant linear mode. Therefore, linear 
theory may be used to aid theoretical predictions even in strongly turbulent regimes. 
In the inertial range of solar wind turbulence, most fluctuations display properties consistent with Alfv{\' e}n 
waves (e.g.,~Refs.~\cite{Belcher1971, Bale2005}), thus motivating the use of magnetohydrodynamics (MHD) 
at scales larger than the proton gyroradius. On the other hand, the question regarding the most 
relevant linear modes in the kinetic range of the solar wind has been the subject of some 
controversy~\cite{Howes2008, Shaikh2009, Podesta2010, Podesta2012, 
Verscharen2012, Sahraoui2012, Boldyrev2013, Chen2013, Narita2016}. 
The leading two wavelike models of kinetic-scale turbulence are presently the kinetic Alfv{\' e}n
wave (KAW) turbulence model \cite{Howes2008, Schekochihin2009, Boldyrev2013} 
and the whistler wave turbulence model \cite{Stawicki2001, Galtier2003, Cho2004, 
Galtier2006, Saito2008, Gary2009}. Upon balancing the linear wave crossing 
time with the nonlinear time, critical balance for both types of 
modes (KAWs and whistlers) predicts an anisotropy given by $k_{\parallel}\propto k_{\perp}^{1/3}$, 
assuming that possible corrections due to intermittency and dissipative effects can 
be neglected \cite{Cho2004, Howes2008, Schekochihin2009}. 
Here, $k_{\parallel}$ should be understood as a characteristic 
wave number along the \emph{local} mean magnetic field~\cite{Cho2000} 
at perpendicular scale $l_{\perp}\sim 1/k_{\perp}$. Observational evidence suggests
that the kinetic-scale fluctuations are predominantly of KAW type~\cite{Bale2005, 
Sahraoui2010, Salem2012, Chen2013}, although there also exists some evidence in support 
of whistler waves~\cite{Perschke2013, Narita2016}.

Complementary to observations and theory, numerical simulations of kinetic-scale solar wind turbulence 
have attracted a great deal of interest~\cite{Howes2008a, Saito2008, 
Cho2009, Verscharen2012, Boldyrev2012, Wu2013a, Karimabadi2013a, 
TenBarge2013a, Chang2014, Vasquez2014, Valentini2014, Servidio2015, Told2015, Wan2015, 
Franci2015, Parashar2016, Gary2016, Kobayashi2017, Cerri2017, Groselj2017}. 
However, capturing the entire range of kinetic physics in
a turbulent simulation has proven difficult due to the immense computational requirements of the problem. For this 
reason, a number of previous works employed various simplifications of the first-principles kinetic description in
three spatial dimensions. These simplifications typically involve various reduced-kinetic approximations
\cite{Howes2008a, TenBarge2013a, Told2015, Servidio2015, Cerri2017} and/or restrictions to a two-dimensional
geometry~\cite{Saito2008, Verscharen2012, Wu2013a, Valentini2014, Servidio2015, Groselj2017}. Only recently
have fully kinetic, three-dimensional (3D) simulations become computationally 
accessible~\cite{Chang2014, Wan2015, Gary2016, Zhdankin2017,Hughes2017}. 
Previous works employing 3D fully kinetic simulations were aimed at 
different aspects such as whistler wave turbulence \cite{Chang2014, Gary2016}, intermittent heating \cite{Wan2015},
particle acceleration in the highly relativistic regime~\cite{Zhdankin2017}, or bulk plasma heating by KAW
turbulence \cite{Hughes2017}. Thus, even though there exists observational evidence for the transition into 
KAW turbulence at kinetic scales~\cite{Bale2005, 
Sahraoui2010, Salem2012, Chen2013}, supplemented by evidence of critical balance in 
gyrokinetic~\cite{TenBarge2012}, electron MHD~\cite{Cho2004, Cho2009}, and Landau fluid simulations~\cite{Sulem2016}, 
the natural occurrence of the transition has to our knowledge never 
been convincingly demonstrated in a 3D fully kinetic simulation.

In this Letter, we try to fill in a long-standing gap in the literature, and perform a
3D fully kinetic plasma turbulence simulation in order to demonstrate the feasibility of 
the critically balanced KAW turbulence model from first principles. Using a simulation setup 
broadly resembling the typical conditions at the tail of the MHD inertial range and 
at subion scales of the slow solar wind, 
we show that the ratios of the turbulent spectra between ion and electron scales are 
consistent with theoretical expectations for KAWs. Furthermore, we perform a first-time 
direct calculation of the \emph{local} scale-dependent anisotropy in a 3D kinetic 
simulation of sub-ion-scale plasma turbulence. From the anisotropy,
we infer the ratio of linear to nonlinear time scales and obtain an order unity estimate 
in the subion range, suggesting that the kinetic cascade is close to a state of critical balance.

{\em Simulation details.---}The triply periodic simulation box dimensions in units of the ion inertial length $d_i$ are $L_{\perp} = 16.97d_i$ and 
$L_z = 42.43d_i$ in directions perpendicular and parallel to the mean magnetic field $\vec B_0 = B_0\hat{\vec e}_z$,
respectively. The initial condition is similar to the one used in Ref.~\cite{Li2016} and consists of 
counterpropagating Alfv\' en waves with wave numbers $(k_{\perp,0}, 0, \pm k_{z,0})$, $(0, k_{\perp,0}, \pm k_{z,0})$, 
and $(2k_{\perp,0}, 0, \pm k_{z,0})$,  where 
$k_{\perp,0} = 2\pi/L_{\perp}$ and $k_{z,0} = 2\pi/L_z$. A different phase is 
used for each mode. ``Alfv\' en waves'' are to be understood here 
in the usual sense of MHD with corresponding 
perpendicular fluid velocity $\delta\vec u_{\perp}$ and 
magnetic field $\delta\vec B_{\perp}$ perturbations. Each pair of counterpropagating waves has equal 
amplitudes, such that the mean cross-helicity $H_c = \langle\delta\vec u\cdot\delta\vec B\rangle$ is zero
(results from a second simulation with nonvanishing cross-helicity are included in 
Supplemental Material~\footnote{See Supplemental Material at \url{http://link.aps.org/supplemental/10.1103/PhysRevLett.120.105101},
which includes Refs.~\cite{Lithwick2007,Beresnyak2008}, for an 
animation of the turbulent dynamics and for results from a second simulation 
with nonvanishing mean cross-helicity.}). Ions and electrons have an initial Maxwellian velocity 
distribution with equal temperatures $T_0$ and uniform 
densities $n_0$, corresponding to a thermal velocity $v_{{\rm th},i} = \sqrt{2T_0/m_i} = 0.031c$ 
for ions and $v_{{\rm th},e} = \sqrt{2T_0/m_e} = 0.25c$ for electrons~\footnote{A recent study~\cite{Groselj2017}, 
using $v_{{\rm th},i}=0.019c$ and $v_{{\rm th},e}=0.19c$, found excellent agreement between two-dimensional (relativistic) 
fully kinetic and (nonrelativistic) gyrokinetic turbulence simulations at sub-ion scales for $\beta_i=0.5$. 
Similarly, a study of the whistler anisotropy instability~\cite{Hughes2016}, using $0.035c \leq v_{{\rm th},e} \leq 0.14c$, 
found no clear physical dependence on $v_{{\rm th},e}$ for a fixed $\beta_e$. Thus, we do not expect significant 
modifications of our results due to artificially large thermal velocities.}, where $c$ is the light speed, $m_i$ is the 
ion mass, and $m_e$ is the electron mass. We also 
initialize a self-consistent electric current according to $\vec J = (c/4\pi)\nabla\times\delta\vec B_{\perp}$.
A reduced ion-electron mass ratio of $m_i/m_e = 64$ is used and 
the electron plasma to cyclotron frequency ratio is $\omega_{pe}/\Omega_{ce} = 2.83$. 
The ion plasma beta is $\beta_i = 8\pi n_0 T_0/B_0^2 = 0.5$. The initial turbulence amplitude $\epsilon = \delta B/B_0 = 
\delta u/v_A$, where $v_A=B_0/\sqrt{4\pi n_0m_i}$ is the Alfv\' en speed, $\delta u$ is the root-mean-square 
fluid velocity, and $\delta B$ is the root-mean-square fluctuating magnetic field, 
is chosen such as to satisfy the critical balance condition ($k_{\perp}\delta B = k_{\parallel}B_0$) 
at the box scale: $\epsilon = L_{\perp}/L_z = 0.4$. 
The physical setup resembles the plasma conditions inferred from solar wind 
measurements \cite{Bale2005, Wicks2010, Chen2010, Chen2016} 
in the following ways: (i) an anisotropy is imposed at the box scale ($k_{z,0} < k_{\perp,0}$),
(ii) the initial condition consists of counterpropagating, oblique Alfv{\' e}n waves, (iii) the initial turbulence amplitude 
is chosen such as to satisfy critical balance, and (iv) the plasma parameters are similar to those 
typically found in the solar wind (plasma beta and ion-electron temperature ratio both of order unity).

We perform the simulation using the 
particle-in-cell code \textsc{osiris} \cite{Fonseca2002, Fonseca2008}. The spatial
resolution is $(N_x,N_y,N_z)=(768,768,1536)$. We employ on average 64 particles per cell per species. 
The charge distribution of each finite-size particle is represented by  
third-order cubic splines \cite{Birdsall2005}, which improve energy conservation and reduce the relative amount 
of particle noise compared to lower-order splines \cite{Cormier-Michel2008, Munoz2014}. At each step, 
we also apply a second-order compensated binomial filter \cite{Birdsall2005} 
on the electric current and on the electromagnetic fields felt by the particles. The total energy increase 
due to numerical heating is kept below 0.033\% during the entire simulation.
To reduce particle noise, the data used for the spectral and 
scale-dependent anisotropy analysis is short-time averaged 
over a time window of duration $\Delta t = 2.4\Omega_{ce}^{-1}$, 
where $\Omega_{ce}=e_0B_0/(m_ec)$ and $e_0$ is the elementary charge.

\begin{figure}[ht]
\centering
\includegraphics{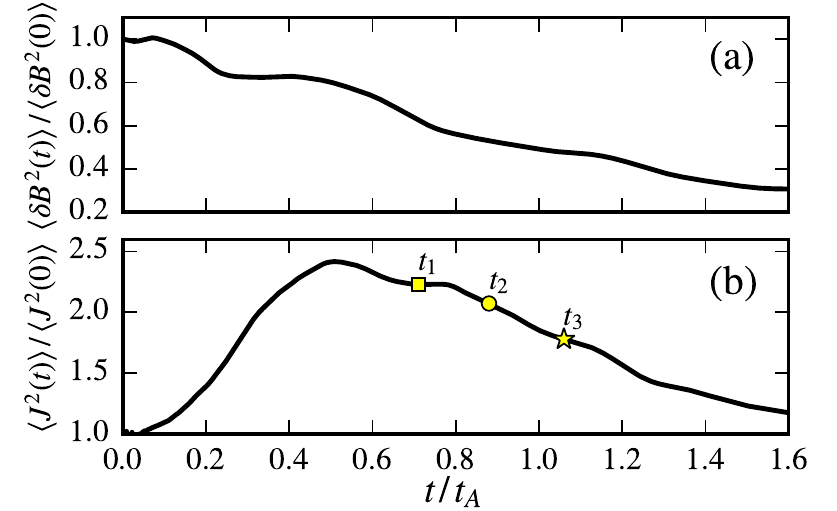}
\caption{Time evolution of the mean magnetic energy (a) and of the 
mean-square electric current (b). The curves are normalized to the values at $t=0$. 
The markers in panel (b) denote the times at which we 
analyze the spectral properties ($t_1/t_A = 0.71, t_2/t_A=0.88, t_3/t_A = 1.06$).\label{pic:b_and_j_trace}}
\end{figure}

{\em Global evolution.---}The global evolution during the turbulent decay is 
illustrated in Fig.~\ref{pic:b_and_j_trace} by plotting the mean fluctuating magnetic energy and the mean-square 
electric current versus time. We take the box-scale Alfv{\' e}n transit time, $t_A=L_z/v_A$, as the basic 
time unit. The markers in Fig.~\hyperref[pic:b_and_j_trace]{\ref*{pic:b_and_j_trace}(b)} are used to 
indicate the times at which we analyze the turbulence spectral properties in what follows. 
The magnetic energy decreases throughout the simulation as a result of ion and electron heating.  By the end of the simulation, 
the species internal energy increases by 17\% for ions and by 15\% for electrons (relative to the value at $t=0$), 
whereas the bulk fluid energy decreases by 76\%. On the other hand, the electric current
undergoes an initial transient, during which it is rapidly amplified, before it eventually starts to decrease. 
The rapid current amplification can be attributed to current sheet formation~\cite{Servidio2015, Matthaeus2015}.
Indeed, a visual inspection of the 3D structure of the electric current (not shown here) 
reveals that the turbulent structures are mainly sheetlike 
(see Supplemental Material~\cite{Note1} for an animation, showing how the current sheets form).

\begin{figure}[ht]
\centering
\includegraphics{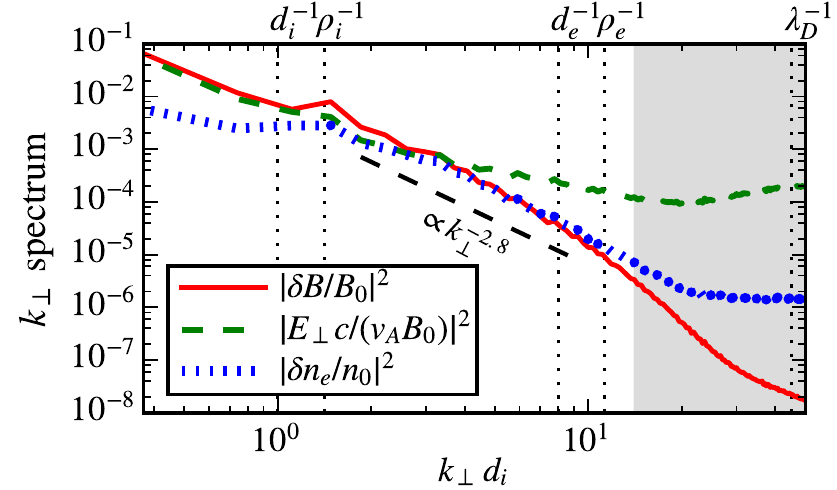}
\caption{One-dimensional $k_{\perp}$ spectra of magnetic, perpendicular electric,
and density fluctuations at time $t_1=0.71 t_A$. 
The $-2.8$ slope is shown for reference. Gray shading is used to indicate the 
range of scales dominated by particle noise.\label{pic:spectrum}}
\end{figure}

{\em Turbulent spectra and spectral ratios.---}We compute 
the one-dimensional perpendicular wave number spectra $E(k_{\perp})$ by summing the squared amplitudes of Fourier modes contained 
in a given perpendicular wave number shell of width $\Delta k_{\perp} = 2\pi/L_{\perp}$, followed by an average along the 
$z$ direction. The shells are nonoverlapping and centered at integer values of $\Delta k_{\perp}$. 
We approximate the perpendicular wave vectors as $\vec k_{\perp}\approx (k_x,k_y)$. That is, the perpendicular direction is
defined with respect to $\vec B_0$~\footnote{The definition is 
consistent with previous works \cite{Cho2000, Maron2001, TenBarge2012} and is motivated 
by the fact that the local $\vec k_{\perp}$ is to lowest order in $\epsilon=\delta B/B_0$ 
equal to ($k_x$, $k_y$), whereas, even to lowest order in $\epsilon$, 
$k_z$ cannot be considered representative 
for the local $k_{\parallel}$~\cite{Maron2001, TenBarge2012}.}.
In Fig.~\ref{pic:spectrum} we show the spectra of the 
magnetic ($\delta\vec B$), perpendicular electric ($\vec E_{\perp}$), and electron density 
fluctuations ($\delta n_e$) at time $t_1=0.71t_A$. 
Similar results are obtained at later stages of the turbulent decay (not shown here) at times
$t_2$ and $t_3$ marked in Fig.~\ref{pic:b_and_j_trace}. Dotted vertical lines are 
used in Fig.~\ref{pic:spectrum} to indicate various kinetic scales: the species inertial
length $d_s = c/\omega_{ps}$, where $\omega_{ps}=\sqrt{4\pi e_0^2n_0/m_s}$ 
and $s=i,e$ is the species index, the species Larmor radius $\rho_s = v_{{\rm th},s}/\Omega_{cs}$, 
where $\Omega_{cs}=e_0B_0/(m_sc)$, and the Debye scale $\lambda_D = v_{{\rm th},e}/(\omega_{pe}\sqrt 2)$.
The sub-ion-scale spectra are in relatively good agreement with a number of observational studies 
\cite{Bale2005, Alexandrova2009, Sahraoui2010, Chen2013}, albeit with some limitations due to the reduced 
ion-electron mass ratio in our simulation. 
In particular, the local slope of the magnetic 
energy spectrum is consistent with the typical values of spectral exponents 
observed in the solar wind \cite{Alexandrova2009, Sahraoui2010}, 
even though a well-defined sub-ion-scale power law cannot be established.
The lack of a well-defined power law, should one in fact exist, can presumably 
be attributed to the proximity of electron kinetic scales, which may cause a steepening of the 
spectral slope due to collisionless damping via the electron Landau resonance~\cite{Howes2008,TenBarge2013a,Told2015,
Kobayashi2017,Groselj2017}. Indeed, solar wind observations \cite{Alexandrova2009} and gyrokinetic simulations 
\cite{TenBarge2013a, Told2015} with realistic proton-electron mass ratios show a steepening of the 
magnetic energy spectra as the wave number approaches the electron scales. 

Looking at the results for $\delta n_e$ and $\vec E_{\perp}$, we find that the electric field spectrum flattens in the 
kinetic range and separates from the magnetic energy, whereas the density spectrum converges toward a 
near equipartition with the magnetic spectrum in appropriately normalized units \cite{Boldyrev2013,Chen2013}. Both 
of these features are in agreement with solar wind observations~\cite{Bale2005,Chen2013}. Most importantly, the 
near equipartition among density and magnetic fluctuations in the subion range is a key property of KAWs, as
opposed to the weakly compressible ($(|\delta n_e|/n_0)^2\ll (|\delta B|/B_0)^2$) 
whistler waves~\cite{Gary2009,Boldyrev2013,Chen2013}. In the 
asymptotic limit
\begin{align}
&& 1/\rho_i \ll k_{\perp} \ll 1/\rho_e, && k_{\parallel}\ll k_{\perp}, &&
\label{eq:limit}
\end{align}
assuming singly charged ions, and equal ion and electron temperatures,
the analytical prediction for KAWs reads~\cite{Boldyrev2013} $(\beta_i + 2\beta_i^2)(|\delta n_e|/n_0)^2 \sim (|\delta B|/B_0)^2$. 
Thus, for $\beta_i=0.5$ we have $(|\delta n_e|/n_0)^2 \sim (|\delta B|/B_0)^2$, 
in agreement with our results presented in Fig.~\ref{pic:spectrum}. The difference between the density 
and magnetic energy spectral slopes seen in Fig.~\ref{pic:spectrum} is a trend 
not captured by the asymptotic prediction. It is, however, fully consistent 
with results from nonlinear gyrokinetic simulations~\cite{Groselj2017}.

\begin{figure}[ht]
\centering
\includegraphics{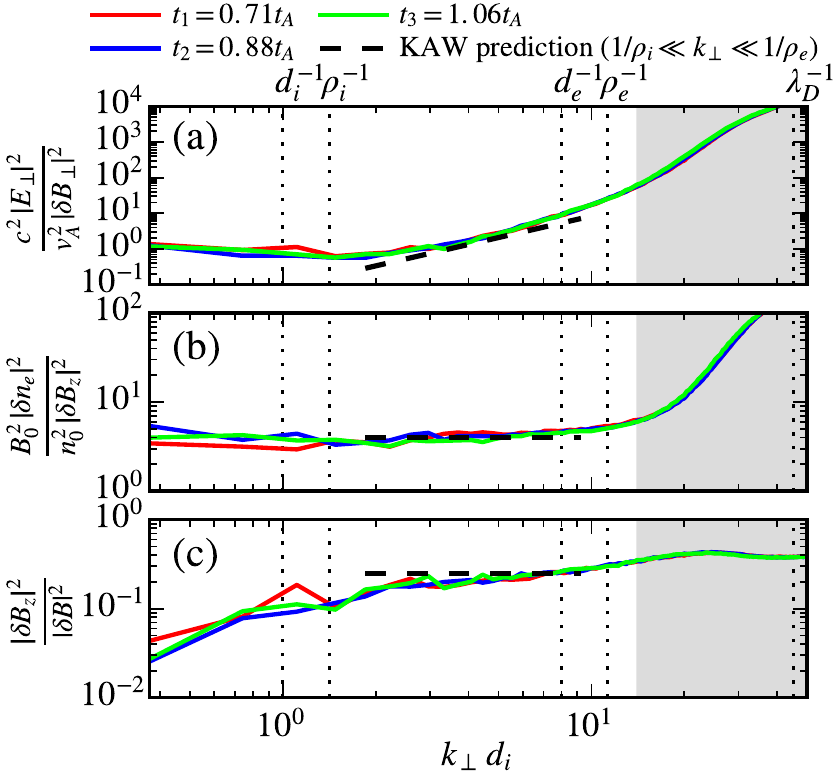}
\caption{Ratios of the $k_{\perp}$ spectra obtained from the 
simulation (solid lines; see text for further details). Dashed lines show the analytical predictions 
for KAWs~\cite{Schekochihin2009, Boldyrev2013}.\label{pic:spec_ratios}}
\end{figure}

To further demonstrate that the sub-ion-scale fluctuations are consistent with theoretical expectations for KAWs, we 
consider the following ratios of the one-dimensional spectra~\cite{Gary2009, Schekochihin2009, 
Boldyrev2013, Groselj2017}:
\begin{align}
\frac{(|E_{\perp}|c/v_A)^2}{|\delta B_{\perp}|^2}& \sim \frac{(k_{\perp}\rho_i)^2}{4+4\beta_i},& \nonumber
\frac{(|\delta n_e|/n_0)^2}{(|\delta B_{\parallel}|/B_0)^2} & \sim \frac{1}{\beta_i^2}, \\
\frac{|\delta B_{\parallel}|^2}{|\delta B|^2} & \sim \frac{\beta_i}{1+2\beta_i}. &
\label{eq:ratios}
\end{align}
The above expressions are obtained from linearized kinetic equations in the limit~\eqref{eq:limit} for singly charged ions, 
and equal ion and electron temperatures. The turbulence spectral ratios 
are compared against the analytical predictions in Fig.~\ref{pic:spec_ratios}~\footnote{Since 
the predictions involving the parallel field $\delta B_{\parallel}$ are independent of $\vec k$ 
and are not expected to give more than just the correct order of magnitude, we approximate the spectral 
ratios using $\delta B_z$ instead of $\delta B_{\parallel}$.}.
Good agreement between the linear KAW theory and the simulation is found for all ratios. The 
results are also in good agreement with nonlinear gyrokinetic simulations~\cite{Groselj2017}.
Considering the fact that the initial fluctuation amplitude in our simulation is relatively large, our simulation box is 
only moderately elongated along $z$, and the ion-electron mass ratio has been reduced, the agreement with 
theoretical predictions is quite remarkable and indicates a certain robustness of the KAW cascade, 
beyond the limits of gyrokinetic theory, in the context of which 
KAW turbulence has most frequently been studied~\cite{Howes2008a,TenBarge2013a,Told2015,BanonNavarro2016}.

\begin{figure}[ht]
\centering
\includegraphics{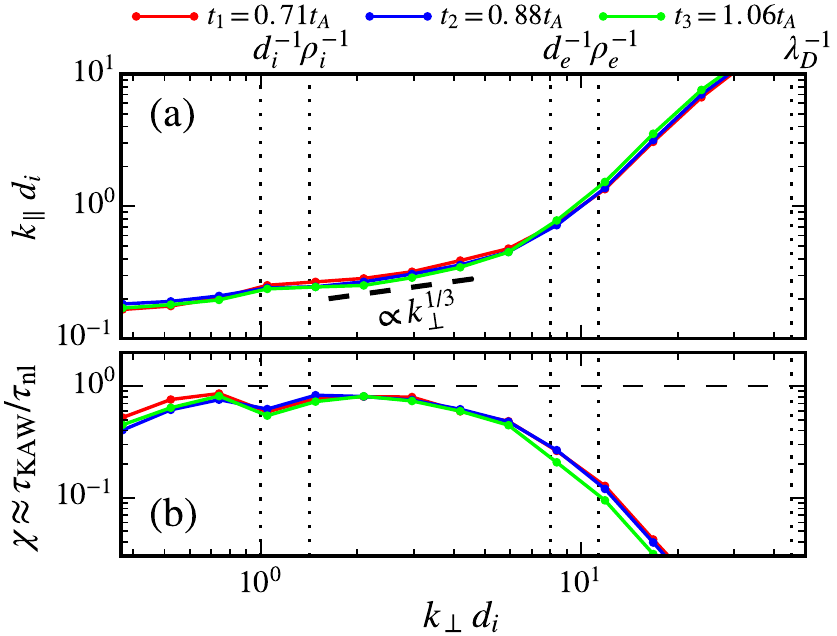}
\caption{Scale-dependent anisotropy with respect to the direction 
of the \emph{local} mean magnetic field (a) and the scale-dependent 
ratio of the linear (KAW) and nonlinear time scales (b). The 
1/3 slope in panel (a) is shown for reference. \label{pic:anisotropy}}
\end{figure}

{\em Scale-dependent anisotropy.---}Finally, we consider the local scale-dependent anisotropy of the kinetic 
turbulence. We employ the method introduced by Cho and Lazarian~\cite{Cho2004,Cho2009}, 
which we summarize here briefly as follows. At a given perpendicular wave number
$k_{\perp}$, we define a local mean magnetic field $\vec B_{0,k_{\perp}}$ and a local fluctuating field $\delta\vec B_{k_{\perp}}$.
The local mean field is obtained by eliminating the Fourier modes with perpendicular wave numbers greater than $k_{\perp}/2$
and the fluctuating field is obtained by eliminating the modes with wave numbers less than $k_{\perp}/2$ or 
greater than $2k_{\perp}$. The characteristic local parallel 
wave number $k_{\parallel}$ at scale $l_{\perp}\sim 1/k_{\perp}$ is then approximated as~\cite{Cho2009}
\begin{equation}
k_{\parallel} \approx \left(\frac{\bigl\langle\left|\vec B_{0,k_{\perp}}\cdot\nabla \delta\vec B_{k_{\perp}}\right|^2\bigr\rangle}
{\langle B_{0,k_{\perp}}^{2}\rangle\langle\delta B_{k_{\perp}}^2\rangle}\right)^{1/2},
\end{equation}
where $\langle\dots\rangle$ represents a space average. In addition, we estimate 
the nonlinearity parameter 
$\chi \approx  k_{\perp}\langle\delta B_{\perp,k_{\perp}}^2\rangle^{1/2} / \bigl(k_{\parallel}B_0\bigr)$~\cite{Cho2004,Howes2008,Schekochihin2009}, 
which can be regarded as an approximation for the ratio of linear (KAW) and nonlinear time scales. For a 
critically balanced cascade, it is expected by definition that $\chi\sim 1$. The results are shown in Fig.~\ref{pic:anisotropy}.
Over a limited subion range, the anisotropy scaling is broadly consistent with the 
standard critical balance prediction, $k_{\parallel} \propto k_{\perp}^{1/3}$~\cite{Howes2008,Schekochihin2009}, although the
scale separation in the simulation is to small to determine the scaling precisely. The estimated
nonlinearity parameter is order unity at subion scales and exhibits a weak dependence on $k_{\perp}$. The 
scale dependence of $\chi$ could be possibly attributed to dissipative effects and/or intermittency~\cite{Boldyrev2012}. 
Moreover, supposing linear modes other than KAWs are energetically significant, 
they could bias the anisotropy estimation of the KAW portion of the 
cascade. Within the limits of the spectral ratios 
analysis, we do not find evidence for the latter possibility. We also confirmed that the 
anisotropy does not change significantly upon inclusion of a moderate 
mean cross-helicity (see Supplemental Material~\citep{Note1}). The question whether or not our conclusions 
are influenced by the reduced ion-electron mass ratio of 64 or by the lack 
of an external turbulence forcing is left for future studies. 
Nonetheless, the local scale-dependent anisotropy calculation performed in this work provides 
the first reference values obtained from a 3D fully kinetic simulation of KAW turbulence.

{\em Discussion and conclusions.---}This Letter presents a 3D fully kinetic simulation of plasma turbulence 
under conditions relevant to the solar wind. 
We show that the spectral properties in the subion range are 
consistent with theoretical expectations for KAWs. The initial perturbations at the start of 
the simulation are restricted to scales above the ion inertial length. Furthermore, 
the initially excited Alfv\' en waves are only moderately oblique. 
Therefore, it is not obvious from a theoretical perspective that kinetic Alfv\' en fluctuations
should dominate at subion scales. Other possibilities, such as whistler wave turbulence, cannot be ruled out.
However, that is not what we observe. A direct calculation of the local scale-dependent anisotropy
is also performed. This allows for an estimate of the nonlinearity parameter $\chi$, which is in broad agreement
with critical balance ($\chi\sim 1$) at subion scales~\cite{Howes2008,Schekochihin2009}.

Our work has important implications for the fundamental understanding of 
kinetic turbulence in weakly collisional plasmas, such as the solar wind, where a number of experimental 
studies already support the KAW turbulence scenario~\cite{Bale2005, Sahraoui2010, Salem2012, Chen2013}. 
Several alternatives or extensions of the KAW turbulence theory have been considered, such as a transition to
whistler turbulence deep in the subion range~\cite{Shaikh2009,Podesta2010}, 
or reconnection-mediated kinetic turbulence~\cite{Cerri2017a,Mallet2017,Loureiro2017,Franci2017}. Given that 
our simulation covers only a moderate range of scales, it is presently difficult to assess the hypothetical role of these features and a 
definitive answer is left for future works. In this Letter we demonstrated that, even when the full range of 3D kinetic physics 
is retained, the phenomenology of critically balanced KAW turbulence remains highly relevant. 
Thus, the KAW turbulence theory seems to provide at least a reasonable starting point, upon which more refined models could be built.

\begin{acknowledgments}

{\em Acknowledgements.---}We gratefully acknowledge helpful conversations with A.~\mbox{Ba{\~ n}{\' o}n} \mbox{Navarro}, D.~\mbox{Told}, 
and C.H.K.~\mbox{Chen}. We also thank S.S.~Cerri for comments on the manuscript. 
Computing resources were provided by the Gauss Centre 
for Supercomputing/Leibniz Supercomputing Centre under Grant No.~pr74vi, 
for which we also acknowledge principal investigator J.~B\" uchner for providing access 
to the resource. N.F.L.~was supported by the National Science Foundation
(NSF) CAREER Grant No.~PHY-1654168.
The authors would like to acknowledge the OSIRIS Consortium, consisting of 
UCLA and IST (Lisbon, Portugal) for the use of \textsc{osiris} and for providing access to the \textsc{osiris} framework.
D.G. thanks F.~\mbox{Tsung}, V.~\mbox{Decyk}, and W.~\mbox{Mori} for helpful discussions about 
the particle-in-cell method and simulations with the \textsc{osiris} code.

\end{acknowledgments}



%


\end{document}